\documentclass[journal,twoside,twocolumn]{IEEEtran}

\usepackage{cite}       

\usepackage{graphicx}   

\usepackage{psfrag}    

\usepackage{subfigure} 

\usepackage{amsmath}   

\usepackage{multirow}   

\usepackage{upgreek}    

\usepackage{amsmath}

\usepackage{amssymb}

\usepackage{amsthm}

\usepackage{mathrsfs}

\usepackage{algpseudocode}

\usepackage{algorithm}

\usepackage{bm}

\usepackage{color}

\usepackage{balance}   

\usepackage[colorlinks=false, linkcolor=blue, bookmarks=false]{hyperref}

\newcommand{\BLUE}[1]{\textcolor{black}{{#1}}}

\begin{document}
\title{\LARGE{A Preamble Collision Resolution Scheme via Tagged Preambles\\ for Cellular IoT/M2M Communications}}
\author{Han~Seung~Jang,~\IEEEmembership{Student Member,~IEEE}, Su Min Kim,~\IEEEmembership{Member,~IEEE}
\\Hong-Shik~Park,~\IEEEmembership{Member,~IEEE}, and Dan~Keun~Sung,~\IEEEmembership{Fellow,~IEEE}
\thanks{H. S. Jang, H. S. Park, and D. K. Sung are with the School of Electrical Engineering, College of Engineering, KAIST, Daejeon 34141, Korea (e-mail: \{jhans, park1507, dksung\}@kaist.ac.kr).}
\thanks{S. M. Kim is with the Department of Electronics Engineering, Korea Polytechnic University, Siheung 15073, Korea (e-mail: suminkim@kpu.ac.kr).}
\thanks{This work has been submitted to the IEEE for possible publication. Copyright may be transferred without notice, after which this version may no longer be accessible.}
}
%

\maketitle
%
\begin{abstract}
In this paper, we propose a preamble (PA) collision resolution (PACR) scheme based on multiple timing advance (TA) values captured via tagged PAs.
In the proposed PACR scheme, tags are embedded in random access (RA) PAs and multiple TA values are captured for a single detected PA during a tag detection procedure.
The proposed PACR scheme significantly improves RA success probability for stationary machine nodes since the nodes using collided PAs can successfully complete the corresponding RAs using exclusive data resource blocks.
\end{abstract}
\section{Introduction}
\IEEEPARstart{C}{ellular} random access (RA) is one of latest active research areas in Internet of Things (IoT)/machine-to-machine (M2M) communications \cite{ref_Laya, ref_Islam} since the RA is a mandatory preceding procedure for communications between a node and an eNodeB in cellular networks.
Thus far, most studies have focused on how to renovate or innovate the current cellular RA system by reflecting main features of IoT/M2M communications such as a massive number of nodes, a variety of data patterns and sizes, and a wide range of access priorities.

Especially, RA overload problems caused by a limited amount of resource in preambles (PAs) and physical uplink shared channels (PUSCHs) have been highly concerned \cite{ref_Osti}.
The approaches to resolve the RA overload problems can be categorized into PA collision mitigation \cite{ref_Jang}, PUSCH resource collision avoidance \cite{ref_Jang5}, PUSCH resource reuse mechanism \cite{ref_Jang4,ref_Liang}, and access class barring (ACB) \cite{ref_Overload2,ref_ACB1}.
\BLUE{Various RA schemes for access prioritization were proposed based on pricing-based load control \cite{ref_koseoglu2017pricing} and PA allocation \cite{ref_vilgelm2017latmapa}.}
However, only a few studies considered the feature of stationary machine nodes in cellular RA even if \BLUE{there exist a lot of application scenarios deploying stationary machine nodes such as sensors and smart meters.} In principle, timing advance (TA) values are delivered via RA response (RAR) messages at the second step of RA procedure so that a node adjusts its uplink timing synchronization.
The TA value of a stationary node is generally fixed and invariant, and it can be already known to the node since the first RA.
This feature can be utilized for avoiding PUSCH resource collisions \cite{ref_Ko} and optimizing ACB parameters \cite{ref_Wang}.

In the conventional RA scheme, PA collisions, which occur when multiple nodes choose an identical PA, cause subsequent PUSCH resource collisions due to late PA collision detection following after detection of PUSCH resource collisions at the third step of RA procedure.
\BLUE{Recently, an early PA collision detection (e-PACD) scheme based on tagged PAs has been proposed \cite{ref_Jang5}. It can fundamentally eliminate PUSCH resource collisions at the third step of the RA procedure and reduce the RA delay. However, the e-PACD scheme cannot eventually resolve the preamble collisions.}
\BLUE{Liang \textit{et al.} \cite{ref_Liang} proposed a non-orthogonal RA scheme, which can partially resolve PA collisions and utilize non-orthogonal PUSCH resources by a collided user group. However, this scheme is only applicable when two collided nodes transmitting the identical PA are sufficiently apart from each other.}

In this paper, we further utilize the tagged PAs in order to resolve PA collisions fundamentally. More specifically, multiple TA values are captured by the tagged PAs at the receiver and they are used for the collision resolution. The proposed PA collision resolution (PACR) scheme enables a stationary node with a collided PA to use an exclusive PUSCH resource at the third step by recognizing not only the used PA index but also its own TA value included in an RAR message at the second step.
The performance of the proposed PACR scheme is analyzed and evaluated in terms of TA capturing accuracy at PHY layer and RA success probability at MAC layer.
The proposed PACR scheme significantly improves the RA success probability of stationary nodes, which occupy a significant portion of a huge population in cellular IoT/M2M networks, since even the nodes using collided PAs can successfully complete the corresponding RA using exclusive PUSCH resource blocks at the third step.
\BLUE{The contributions of this paper are summarized as follows:
\begin{itemize}
			\item We propose a novel way to capture multiple TA values for a single detected PA and a novel PUSCH resource allocation method using the TA values included in RAR messages in order to fundamentally resolve PA collisions.
		\item We analyze the TA capturing accuracy at PHY layer, and the RA success probability and the PUSCH resource collision probability at MAC layer.
		\item The proposed PACR scheme achieves RA success probabilities higher than 90\% and PUSCH collision probabilities lower than 2\%, compared with the conventional RA scheme \cite{ref_LTE} and the Liang's RA schemes \cite{ref_Liang} even in a significantly heavy RA traffic environment.
\end{itemize}
}

\section{Tagged Preambles}
According to \cite{ref_Jang5}, a tagged PA consists of a PA Zadoff-Chu (ZC) sequence and a tag ZC sequence, which are transmitted as a mixed sequence:
\begin{equation}
X_{r,k_i}^{i,l}[n]=\beta(p_{r}^{i}[n]+q_{k_i}^{l}[n]),
\end{equation}
where $\beta$ denotes the signal strength of the tagged PA, and $p_{r}^{i}[n]$ and $q_{k_i}^{l}[n]$ denote the PA sequence and the tag sequence, respectively.
The PA sequence is expressed as $p_{r}^{i}[n]=z_r[(n+iN_{\mathrm{CS}})\bmod{N_{\mathrm{ZC}}}]$ for $n=0,\ldots,(N_{\mathrm{ZC}}-1)$, where $r$ denotes the PA root number, $i$ denotes the randomly selected PA index in $\{0,\ldots,(N_{\mathrm{PA}}-1)\}$, $N_{\mathrm{CS}}$ denotes the cyclic shift size, and $N_{\mathrm{PA}}$ denotes the number of available PAs.
Similarly, the tag sequence is expressed as
$q_{k_i}^{l}[n]=z_{k_i}[(n+lN_{\mathrm{CS}})\bmod{N_{\mathrm{ZC}}}]$ for $n=0,\ldots,(N_{\mathrm{ZC}}-1)$, where $k_i$ denotes the tag root number of the $i$-th PA index, $l$ denotes the randomly selected tag index in $\{0,\ldots,(N_{\mathrm{tag}}-1)\}$, and $N_{\mathrm{tag}}$ denotes the number of available tags.
The tag root number $k_i$ is determined by a tag root mapping function of the PA index $i$, i.e., $k_i=f(i)$, which is exclusive of the PA root number $r$, i.e., $r\neq k_i$, to maintain a cross-correlation property \cite{ref_LTE}.
\BLUE{An expense for sending additional tag sequences in the proposed PACR scheme may be a negative effect on PA detection performance. However, it has been examined that the additional tag sequences do not significantly affect the PA detection probability \cite{ref_Jang5}.}

\vspace*{-2mm}
\section{System Model}
Let $M_i$ denote the number of RA-attempting nodes using the $i$-th PA index for $i=0,\ldots,(N_{\mathrm{PA}}-1)$, and $M=\sum_{i=0}^{(N_{\mathrm{PA}}-1)}M_i$ represents the total number of RA-attempting nodes on the same PRACH.
Let us define the following sets:
\begin{itemize}
\item $\mathcal{I}=\{i_m|m=1,\ldots,M\}$\\
: A set of transmitted PA indices from $M$ nodes.
\item $\mathcal{L}=\{l_m|m=1,\ldots,M\}$\\
: A set of transmitted tag indices from $M$ nodes.
\item $\mathcal{K}=f(\mathcal{I})=\{k_{i_m}=f(i_m)|m=1,\ldots,M\}$\\
: A set of used tag root numbers from $M$ nodes.
\end{itemize}
The $M$ nodes simultaneously transmit their own tagged PAs
\begin{equation}
X_{r,k_{i_m}}^{i_m,l_m}[n]=\beta_{m}(p_{r}^{i_m}[n]+q_{k_{i_m}}^{l_m}[n]),~m=1,\ldots,M,
\end{equation}
where $p_{r}^{i_m}[n]$ and $q_{k_{i_m}}^{l_m}[n]$ denote the PA and tag sequences, and $\beta_{m}$, $r$, $k_{i_m}$, $i_m$, and $l_m$ denote the transmitted power, the PA root number, the tag root number, the PA index, and the tag index for the $m$-th node, respectively.
Then, the received sequence at the eNodeB is expressed as
\begin{equation}
Y_{r,\mathcal{K}}^{\mathcal{I},\mathcal{L}}[n]=\sum_{m=1}^{M}\sum_{e=1}^{E_m}h^e_mX_{r,k_{i_m}}^{i_m,l_m}[(n+t^e_m)\bmod{N_{\mathrm{ZC}}}]+W[n],
\label{eq_received}
\end{equation}
where $h^e_m$ and $t^e_m$ denote the channel coefficient and the delay shift of the $m$-th node for the $e$-th multi-path,  respectively, $E_m$ is the total number of multi-paths for the $m$-th node, and $W[n]$ represents the circular symmetry complex Gaussian noise with variance $\sigma^2$, i.e., $W[n]=W_{\mathrm{R}}[n]+jW_{\mathrm{I}}[n]\sim\mathcal{CN}(0,\sigma^2)$.
\vspace*{-2mm}
\section{Proposed Preamble Collision Resolution}\label{sec_4}
In this section, we first describe how to capture multiple TA values with respect to a single detected PA.
Then, the two-step procedure of the proposed PACR scheme is presented.

\subsection{Capturing of multiple timing advance (TA) values}\label{sec_4a}
\begin{figure}[t]
\centering
\includegraphics[width=3.0 in]{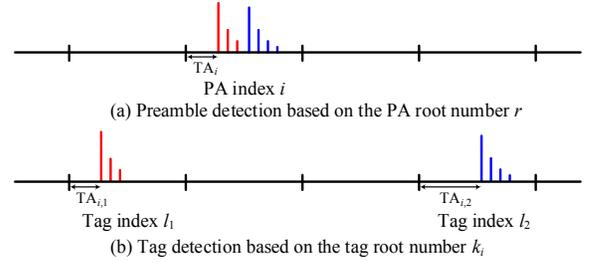}
\vspace*{-3mm}
\caption{An example of capturing two TA values when two nodes use the same PA index $i$ and distinct tag indices $l_1$ and $l_2$.}
\vspace{-3mm}
\label{fig_capturing}
\end{figure}

After the eNodeB receives tagged PAs on PRACH, it first tries to detect PAs using the PA reference sequence with PA root number $r$.
As a result, it obtains a set of detected PA indices $\mathcal{I}^{\prime}=\{i_1,\ldots,i_{J}\}$ where $J$ denotes the number of detected PA indices.
After that, the eNodeB tries to detect the corresponding tags for each of the detected PA indices $i_j\in \mathcal{I}^{\prime}$ based on the tag reference sequence with tag root number $k_{i_j}=f(i_j)$.
During the tag detection for the detected PA $i_j$, the eNodeB may capture $T_{i_j}$ TA values on each of $T_{i_j}$ distinct tag detection zones.
Fig. \ref{fig_capturing} shows an example of capturing two TA values.
In the example, two nodes use the same PA index $i$ and distinct tag indices $l_1$ and $l_2$, respectively. In Fig. \ref{fig_capturing} (a), during the PA detection, the $i$-th PA can be detected but only a single and shorter TA value of $\mathrm{TA}_{i,1}$ can be captured.
However, in Fig. \ref{fig_capturing} (b), during the tag detection using the tag root number $k_{i}$, two tags can be detected and thus, two TA values $\mathrm{TA}_{i,1}$ and $\mathrm{TA}_{i,2}$ can be captured on two different tag detection zones of $l_1$ and $l_2$, respectively.

After all, using the tagged PAs, the eNodeB can capture multiple TA values for each detected PA. According to the number of captured TA values, $T_i$, for the $i$-th detected PA index, the eNodeB can classify the $i$-th detected PA index as single-, double-, triple- or over-access PAs.
However, majority may correspond to single-access and double-access PAs.

\subsection{Procedure of the proposed preamble collision resolution}\label{sec_4b}
\BLUE{\textbf{(1) RAR message generation}: If the eNodeB successfully obtains $T_i$ TA values for the $i$-th detected PA, it can generate $T_i$ RAR messages, each of which includes the $i$-th PA index, one of $T_i$ TA values, and an uplink resource grant for the third step of the RA procedure. Then, it transmits $T_i$ RAR messages to the nodes using the same PA index $i$.
Note that if the TA values captured on different tag detection zones are identical, the eNodeB does not generate an RAR message including this TA value so that the nodes with the identical TA value and PA index can avoid PUSCH resource collisions at the next step and then quickly reattempt another RA.}

\BLUE{\textbf{(2) Finding their own RAR messages}: Since most stationary nodes are aware of their TA values at network-entry instant, they can obtain their own RAR messages by verifying their PA indices and TA values at the second step of the RA procedure. If they decode their own RAR messages, they send RA-step3 data at the third step of the RA procedure on PUSCH informed through their uplink resource grants indicated in the RAR messages. As a result, even though multiple nodes choose the same PA (i.e., resulting in a PA collision in the conventional RA scheme), they can succeed the RA at the third step by using distinct PUSCH resource blocks.}

\begin{figure}[t]
\centering
\includegraphics[width=3.5 in]{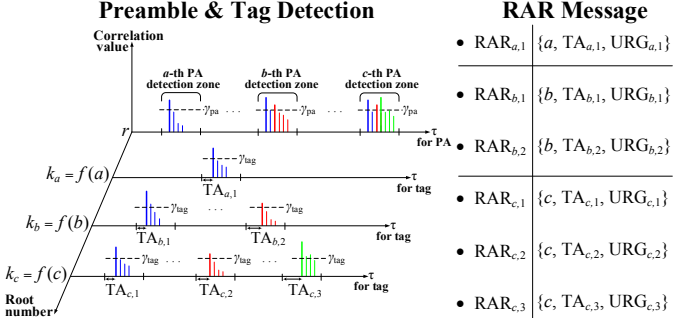}
\vspace{-3mm}
\caption{An illustrative example of the proposed PACR scheme. Single-access PA index: $a$, double-access PA index: $b$, and triple-access PA index: $c$.}
\vspace{-3mm}
\label{fig_main}
\end{figure}

Fig. \ref{fig_main} shows an illustrative example. Here, six nodes in total simultaneously attempt RAs on the same PRACH; Node~1 uses PA index $a$ (a single-access PA); Nodes~2 and 3 use PA index $b$ (double-access PA); Nodes~4, 5, and 6 use PA index $c$ (triple-access PA). For tag indices, we assume that $l_2\neq l_3$ and $l_4\neq l_5\neq l_6$.
The PA indices $a$, $b$, and $c$ can be detected during the PA detection procedure. Then, using the tag root numbers  $k_i=f(i)$ for $i\in\{a,b,c\}$, a single or multiple TA values can be captured with respect to the $i$-th detected PA. The $i$-th detected PA has TA values $\mathrm{TA}_{i,n}$, $n\in\{1,\ldots,T_i\}$, where $T_i$ denotes the number of captured TA values for the $i$-th detected PA. According to the PA detection and TA capturing results, six RAR messages can be generated and each of them includes the detected PA index $i$, the TA value $\mathrm{TA}_{i,n}$, and the uplink resource grant $\mathrm{URG}_{i,n}$ for $i\in\{a,b,c\}$ and $n\in\{1,\ldots,T_i\}$.
\vspace*{0pt}
\section{Performance Analysis}
For analytical tractability and simplicity, we assume that each tagged PA experiences
only a single path and the average channel coefficient is unity as in \cite{ref_Jang5,ref_Jang3}.
Thus, the received sequence in (\ref{eq_received}) is rewritten as
\begin{equation}
Y_{r,\mathcal{K}}^{\mathcal{I},\mathcal{L}}[n]=\sum\nolimits_{m=1}^{M}X^{i_m,l_m}_{r,k_{i_m}}[(n+t_m)\bmod{N_{\mathrm{ZC}}}]+W[n],
\end{equation}
where $r$, $\mathcal{K}$, $k_{i_m}$, and $t_m$ denote the PA root number, the set of used tag root numbers from $M$ nodes, the tag root number of the $m$-th node, and the round-trip delay shift of the $m$-th node, respectively. In the following, we analyze the PA detection probability and the TA capturing accuracy for the $d$-th node using the PA index $i_d$ and the tag index $l_d$, i.e., the tagged PA $X^{i_d,l_d}_{r,k_{i_d}}[n]$. After then, we derive the RA success probability and the PUSCH collision probability at MAC layer.

\vspace*{0pt}
\subsection{Analysis of PA detection probability}
The $i_d$-th PA can be detected by calculating the correlation value $|c_{\{r,\mathcal{K}\},r}[\tau]|$ between the received sequence $Y_{r,\mathcal{K}}^{\mathcal{I},\mathcal{L}}[n]$ and the PA reference sequence $z_{r}[n]$, and then, verifying the $i_d$-th PA detection zone $\mathcal{D}_{i_d}=[i_dN_{\mathrm{CS}},(i_d+1)N_{\mathrm{CS}}-1]$ \cite{ref_Jang5}.
Since there are $M_{i_d}$ nodes using the PA index $i_m=i_d$, $M_{i_d}$ peaks can be detected at each of $\tau\in\bm{\Omega}^{i_d}_{\mathrm{pa}}=\{\omega^{i_d}_j|j=1,\ldots,M_{i_d}\}$, where $\omega^{i_d}_j=\{(i_dN_{\mathrm{CS}}+t_j) \bmod N_{\mathrm{ZC}}\}$.
At time instant $\tau=\omega^{i_d}_j$, $\left|c_{\{r,\mathcal{K}\},r}[\omega^{i_d}_j]\right|$ becomes a random variable (RV) $G^{i_d}_j\sim\mathrm{Rice}(\theta^{i_d}_j,\sigma)$ following a Rician distribution\cite{rice}, whose probability density function (PDF) is expressed as
\begin{equation}
\label{PDF}
\textstyle f_{G^{i_d}_j}(g;\theta^{i_d}_j,\sigma)=\frac{g}{\sigma^2}I_0\left(\frac{g\theta^{i_d}_j}{\sigma^2}\right)\exp\left[-\frac{g^2+(\theta^{i_d}_j)^2}{2\sigma^2}\right], \end{equation}
where $\theta^{i_d}_j=\left|\sqrt{N_{\mathrm{ZC}}}\beta_{j}+\sum_{m=1}^
{M}c_{k_{i_m},r}[\omega^{i_d}_j]\right|$, and $I_0(\cdot)$ is the modified Bessel function of the first kind with the zero-th order.
Therefore, we have $M_{i_d}$ RVs $\{G^{i_d}_j|j=1,\ldots,M_{i_d}\}$ at each of $\bm{\Omega}^{i_d}_{\mathrm{pa}}$ in $\mathcal{D}_{i_d}$.
Let us define the maximum RV $G^{i_d}_{\mathrm{pa}}=\max\{G^{i_d}_j|j=1,\ldots,M_{i_d}\}$.
Finally, the PA detection probability for the PA index $i_d$ is defined as
\begin{align}
\textstyle P^{i_d}_{\mathrm{PA}}(M_{i_d})&\triangleq\mathrm{Pr}[G^{i_d}_{\mathrm{pa}}\geq\gamma_{\mathrm{pa}}]\nonumber\\
&=\textstyle 1-\prod_{j=1}^{M_{i_d}}\left\{1-Q_1\left(\frac{\theta^{i_d}_j}{\sigma},\frac{\gamma_{\mathrm{pa}}}{\sigma}\right)\right\},
\end{align}
where $Q_1(a,b)$ and $\gamma_{\mathrm{pa}}$ denote the Marcum Q function and the PA detection
threshold, respectively.

\vspace*{0pt}
\subsection{Analysis of TA capturing accuracy}
After the PA index $i_d$ is detected, the eNodeB again calculates the correlation value $|c_{\{r,\mathcal{K}\},k_{i_d}}[\tau]|$ between the received sequence $Y_{r,\mathcal{K}}^{\mathcal{I},\mathcal{L}}[n]$ and the tag reference sequence $z_{k_{i_d}}[n]$ using the tag root number $k_{i_d}=f(i_d)$, and verifies each of the tag detection zones $\mathcal{D}_l=[lN_{\mathrm{CS}},(l+1)N_{\mathrm{CS}}-1]$ for $l\in[0, N_{\mathrm{tag}}-1]$.
There are $M_{i_d}$ tags attached to the $i_d$-th PA using $k_{i_m}=k_{i_d}$.
Among them, the tag index $l_d$ of the $d$-th node can be detected on the $l_d$-th tag detection zone $\mathcal{D}_{l_d}=[l_dN_{\mathrm{CS}},(l_d+1)N_{\mathrm{CS}}-1]$ at $\tau=\psi^{i_d}_{l_d}=\{(l_dN_{\mathrm{CS}}+t_d) \bmod N_{\mathrm{ZC}}\}$.
At time instant $\tau=\psi^{i_d}_{l_d}$, $\left|c_{\{r,\mathcal{K}\},k_{i_d}}[\psi^{i_d}_{l_d}]\right|$ becomes an RV $H^{i_d}_{l_d}\sim\mathrm{Rice}(\phi^{i_d}_{l_d},\sigma)$ following a Rician distribution, whose PDF is given in (\ref{PDF}) with a different parameter $\phi^{i_d}_{l_d}=\Big|\sqrt{N_{\mathrm{ZC}}}\beta_{d}+\sum\nolimits_{m=1}^{M}c^{m}_{r,k_{i_d}}[\psi^{i_d}_{l_d}]+\sum\nolimits_{k_{i_m}\neq k_{i_d}}^Mc_{k_{i_m},k_{i_d}}[\psi^{i_d}_{l_d}]\Big|.$

Before $\tau=\psi^{i_d}_{l_d}$, there are $t_d$ instants  $\bm{\Theta}^{i_d}_{l_d}=[l_dN_{\mathrm{CS}}\bmod N_{\mathrm{ZC}},(l_dN_{\mathrm{CS}}+t_d-1)\bmod N_{\mathrm{ZC}}]$ on the tag detection zone $\mathcal{D}_{l_d}$, and at each of them, $|c_{\{r,\mathcal{K}\},k_{i_d}}[\tau]|$ becomes an RV $V^{i_d}_{l_d}[\tau]\sim\mathrm{Rice}(\eta^{i_d}_{l_d}[\tau],\sigma)$ following a Rician distribution, whose PDF is given in (\ref{PDF}) with a different parameter $\eta^{i_d}_{l_d}[\tau]=\Big|\sum\nolimits_{m=1}^{M}c^{m}_{r,k_{i_d}}[\tau]+\sum\nolimits_{k_{i_m}\neq k_{i_d}}^{M}c_{k_{i_m},k_{i_d}}[\tau]\Big|.$

Here, let us define the maximum noise RV $Z^{i_d}_{l_d}\triangleq\max\{V^{i_d}_{l_d}[\tau]|\tau\in\bm{\Theta}^{i_d}_{l_d}\}$, whose cumulative density function (CDF) is expressed as
\begin{equation}\label{tag_noise}
\textstyle \mathrm{Pr}[Z^{i_d}_{l_d}\leq z;\bm{\eta}^{i_d}_{l_d},\sigma]=\prod_{\tau\in\bm{\Theta}^{i_d}_{l_d}}\left\{1-Q_1\left(\frac{\eta^{i_d}_{l_d}[\tau]}{\sigma},\frac{z}{\sigma}\right)\right\},
\end{equation}
where $\bm{\eta}^{i_d}_{l_d}=\{\eta^{i_d}_{l_d}[\tau]|\tau\in\bm{\Theta}^{i_d}_{l_d}\}$.

Finally, we define the TA capturing accuracy for the $d$-th node as the probability that the tag index $l_d$ of the $d$-th node is accurately captured with $t_d$ within the $l_d$-th tag detection zone $\mathcal{D}_{l_d}$.
For that, the RV $H^{i_d}_{l_d}$ at $\tau=\psi^{i_d}_{l_d}$ should be greater than or equal to the tag detection threshold $\gamma_{\mathrm{tag}}$, and the each of $t_d$ noise values in $\bm{\Theta}^{i_d}_{l_d}$ should not exceed $\gamma_{\mathrm{tag}}$.
Therefore, the TA capturing accuracy for the $d$-th node using the $i_d$-th PA index is derived as
\begin{align}\label{eq_TA_capturing}
P_{\mathrm{TA}}^{i_d}(M_{i_d})=\textstyle P^{i_d}_{\mathrm{PA}}(M_{i_d})\mathrm{Pr}[Z^{i_d}_{l_d}< \gamma_{\mathrm{tag}},H^{i_d}_{l_d}\geq \gamma_{\mathrm{tag}}],
\end{align}
where $\mathrm{Pr}[Z^{i_d}_{l_d}<\gamma_{\mathrm{tag}}]$ is given in (\ref{tag_noise}) with $z=\gamma_{\mathrm{tag}}$, and $\mathrm{Pr}[H^{i_d}_{l_d}\geq \gamma_{\mathrm{tag}}]=Q_1(\phi^{i_d}_{l_d}/\sigma,\gamma_{\mathrm{tag}}/\sigma)$.

\vspace*{0pt}
\subsection{Analysis of RA success probability and PUSCH collision probability}
From the viewpoint of MAC layer, we analyze the RA success probability of both the proposed PACR and the conventional RA schemes.
We assume that $M$ stationary machine nodes simultaneously attempt RAs on the same PRACH and a single target node selects a PA index $i$ and a tag index $l$.
Let $N_{\mathrm{PA}}$ and $N_{\mathrm{tag}}$ denote the number of available PAs and the number of available tags, respectively. Here, $N_{\mathrm{tag}}$ is determined by a cell radius $R$ \cite{ref_Jang}.
First, we derive the probability located within a specific TA zone. Let $\epsilon_{\mathrm{TA}}$ denote the distance granularity of a TA value \cite{ref_Wang}.
Within a cell with radius $R$, there exists $N_{\mathrm{TA}}=\lceil R/\epsilon_{\mathrm{TA}}\rceil$ TA zones and the $d$-th TA zone is in the range of $[d\epsilon_{\mathrm{TA}},(d-1)\epsilon_{\mathrm{TA}}]$ for $d\in\{0,\ldots,N_{\mathrm{TA}}-1\}$.
The probability located within the $d$-th TA zone is obtained by
\begin{equation}
P_d=\int_{r_d}^{r_{d+1}}\frac{2r}{R^2}dr=\frac{1}{R^2}\left(r_{d+1}^2-r_{d}^2\right),
\end{equation}
where $r_d=d\epsilon_{\mathrm{TA}}$.

The probability that $a$ nodes choose the same PA index $i$ among total $(M-1)$ stationary nodes (except for the target node) is defined as
\begin{equation}
  \textstyle\mathrm{Pr}[A_i=a,M-1]\triangleq{M-1 \choose a}\left(\frac{1}{N_{\mathrm{PA}}}\right)^a\left(\frac{N_{\mathrm{PA}}-1}{N_{\mathrm{PA}}}\right)^{(M-1-a)},
\end{equation}
where $A_i$ represents an RV for the number of other nodes selecting the same PA index $i$ and ${x \choose y}$ represents the binomial coefficient.
Assuming that PUSCH resources for the third step are sufficient (i.e., the probability of successful PUSCH scheduling is unity), the RA success probability of the proposed PACR scheme for a single stationary node located within the $d$-th TA zone is derived as
\begin{align}
\textstyle&P_{\mathrm{success}}^{\mathrm{prop}}(M,d)\triangleq\mathrm{Pr}[B_{i,l}=1,D_{i,d}=1,M-1] \nonumber\\
\textstyle&=\sum_{a=0}^{M-1}\left\{\left(\frac{N_{\mathrm{tag}}-1}{N_{\mathrm{tag}}}\right)(1-P_d)\right\}^{a}\mathrm{Pr}[A_i=a,M-1],
\end{align}
where $B_{i,l}$ denotes an RV for the number of nodes selecting both the PA index $i$ and the tag index $l$ and $D_{i,d}$ denotes an RV for the number of nodes selecting the PA index $i$ and located within the $d$-th TA zone.
The average RA success probability of the proposed PACR scheme for a single stationary node is determined by
\begin{align}
\textstyle&P_{\mathrm{success}}^{\mathrm{prop}}(M)=\sum\nolimits_{d=0}^{N_{\mathrm{TA}}-1}P_{\mathrm{success}}^{\mathrm{prop}}(M,d) P_d.
\end{align}
Similarly, the RA success probability of the conventional RA scheme is obtained by
\begin{align}
\textstyle P_{\mathrm{success}}^{\mathrm{conv}}(M)\triangleq\mathrm{Pr}[A_i=0,M-1]=\left(\frac{N_{\mathrm{PA}}-1}{N_{\mathrm{PA}}}\right)^{(M-1)}.
\end{align}
Finally, the PUSCH resource collision probabilities of the proposed PACR and the conventional RA schemes are derived, respectively, as
\begin{align}
\textstyle
P_{\mathrm{collision}}^{\mathrm{prop}}(M)&=\textstyle1-\left(\frac{N_{\mathrm{PA}} N_{\mathrm{tag}}-1}{N_{\mathrm{PA}} N_{\mathrm{tag}}}\right)^{(M-1)},\\
P_{\mathrm{collision}}^{\mathrm{conv}}(M)&=1-P_{\mathrm{success}}^{\mathrm{conv}}(M).
\end{align}

\vspace*{-0mm}
\section{Performance Evaluation} \label{sec_5}
\begin{figure}[t]
	\centering
	\includegraphics[width=2.5 in]{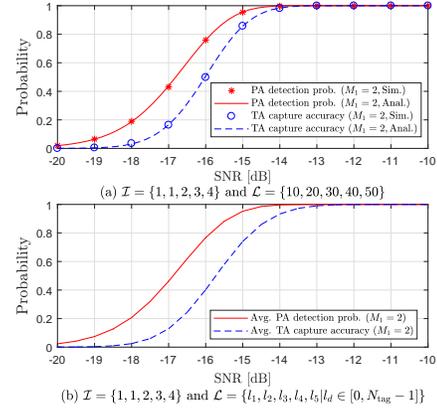}
	\vspace{-3mm}
	\caption{PA detection probability and TA capturing accuracy.}
	\vspace{-4mm}
	\label{fig_pa_detection}
\end{figure}

\begin{table}[t]
    \renewcommand{\arraystretch}{1.2}
    \caption{Simulation parameters and values}
    \label{tbl_parameters}
    \begin{center}
    \vspace*{-5mm}
    \begin{tabular}{c|c}
        \hline
        \hline
        \hspace{-2pt}Parameters\hspace{-2pt}   & Values  \\
        \hline
        \hline
        PA detection threshold, $\gamma_{\mathrm{pa}}$ (dB)         & $-16$\\ \hline
        Tag detection threshold, $\gamma_{\mathrm{tag}}$ (dB)       & $-16$\\ \hline
        Signal-to-noise ratio (SNR), $\beta$ (dB) & $-20$ $\sim$ $-10$ \\ \hline
        Number of RA nodes per PRACH, $M$          & $5$ \\ \hline
         Cell radius, $R$ (km)                   & $0.8, 1.6, 2.4$\\ \hline
         TA Distance granularity, $\epsilon_{\mathrm{TA}}$ (km)                      & $0.8$\\ \hline
        Number of PAs for stationary nodes, $N_{\mathrm{PA}}$                      & $20$\\ \hline
        Number of tags, $N_{\mathrm{tag}}$ (determined by $R$)        & $71$, $51$, $38$\\ \hline
        Number of TA zones, $N_{\mathrm{TA}}$ (determined by $R$)   & $10$, $20$, $30$\\ \hline\hline
        \end{tabular}
    \end{center}
    \vspace*{-6mm}
\end{table}
Table \ref{tbl_parameters} lists the simulation parameter set utilized in simulations.
\BLUE{All simulations were performed by using MATLAB\textsuperscript\textregistered.}
Fig. \ref{fig_pa_detection} shows the PA detection probability and the TA capturing accuracy of the proposed PACR scheme when the number of RA-attempting nodes using PA~1 among five nodes ($M=5$)\footnote{Assuming 50,000 nodes in a cell with an RA arrival rate of 0.5 (one arrival/2 minutes), the corresponding $M$ value becomes approximately 5.} is two ($M_1=2$). In Fig. \ref{fig_pa_detection} (a), the set of transmitted PA and tag indices from five nodes are set to $\mathcal{I}=\{1,1,2,3,4\}$ and $\mathcal{L}=\{10,20,30,40,50\}$, respectively, and the target node with $i_1=1$ and $l_1=10$ has a TA value of $t_1=3$. First of all, the simulation results agree well with the analytical results. The TA capturing accuracy is apart approximately less than $1$ dB from the PA detection probability since the TA capturing accuracy requires the successful PA detection in advance as represented in Eq. (\ref{eq_TA_capturing}).
In Fig. \ref{fig_pa_detection} (b), $\mathcal{L}$ is randomly chosen 1000 times in order to obtain the average values. In this situation, the PA detection probability has the similar value to that in Fig. \ref{fig_pa_detection} (a), but the TA capturing accuracy is apart approximately $0.3$ dB from that in Fig. \ref{fig_pa_detection} (a).

\begin{figure}[t]
	\centering
	\includegraphics[width=2.5 in]{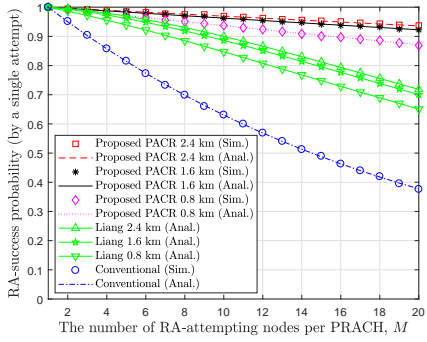}
	\caption{RA success probability for a single RA attempt.}
	\vspace{-4mm}
	\label{fig_ra_success}
\end{figure}

\BLUE{From the viewpoint of MAC layer, we evaluate the proposed PACR scheme in terms of RA success probability and PUSCH resource collision probability, compared with the Liang's RA scheme \cite{ref_Liang} and the conventional RA scheme \cite{ref_LTE}.
The Liang's RA scheme is to partially resolve PA collisions only when signal peaks are sufficiently apart from each other within the same PA detection zone.}
In Fig. \ref{fig_ra_success}, the proposed PACR scheme achieves the RA success probabilities of 93.6\% ($R=2.4$ km), 92.3\% ($R=1.6$ km), and 87.2\% ($R=0.8$ km) for a single RA attempt in a significant RA load situation (e.g. $M=20$), while \BLUE{the Liang's RA scheme achieves the RA success probabilities of 71.8\% ($R=2.4$ km), 70.0\% ($R=1.6$ km), and 65.1\% ($R=0.8$ km)}, and the conventional RA scheme achieves only 37.7\% (the cell radius does not affect the performance in the conventional scheme).
This implies that the proposed PACR scheme can improve the RA success probabilities up to 99.6\% ($R=2.4$ km), 99.4\% ($R=1.6$ km), and 98.4\% ($R=0.8$ km) with two consecutive RA attempts, while \BLUE{the Liang's RA scheme achieves 92.0\% ($R=2.4$ km), 91.0\% ($R=1.6$ km), and 87.8\% ($R=0.8$ km)}, and the conventional RA scheme achieves 61.2\%.
The larger the cell, the higher the RA success probability since the probability that TA values are overlapped is reduced as the cell radius increases.

\begin{figure}[t]
\centering
\includegraphics[width=2.5 in]{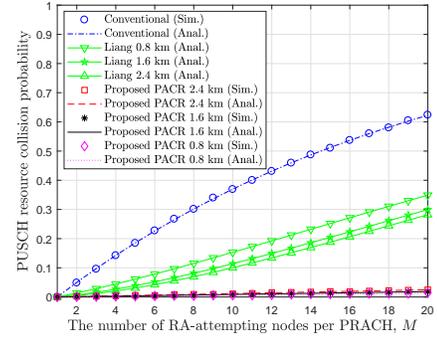}
\caption{PUSCH resource collision probability.}
\vspace{-4mm}
\label{fig_ra_collision}
\end{figure}

In addition to the PA collision resolution capability, even in case of TA overlapped, the proposed PACR scheme is helpful to avoid PUSCH resource collisions since it does not allocate any PUSCH resource for this case. In Fig. \ref{fig_ra_collision}, the proposed PACR scheme achieves significantly lower values in the PUSCH resource collision probability as 1.43\% ($R=0.8$ km), 1.82\% ($R=1.6$ km), and 2.47\% ($R=2.4$ km), compared with 34.9\% ($R=0.8$ km), 29.9\% ($R=1.6$ km), and 28.2\% ($R=2.4$ km) of the Liang's RA scheme and 37\% of the conventional RA scheme when $M=20$.

\section{Conclusion}\label{sec_6}
In this paper, we proposed a PA collision resolution (PACR) scheme which significantly improves RA success probability and reduces PUSCH resource collisions based on the multiple TA values captured via tagged PAs.
The performance of the proposed PACR scheme was mathematically analyzed in terms of the TA capturing accuracy at PHY layer, and the RA success and the PUSCH resource collision probabilities at MAC layer.
Through simulations, it was shown that the proposed PACR scheme is very effective for stationary machine nodes which occupy a significant portion of huge population in future cellular IoT/M2M networks.




%
\balance
\bibliographystyle{./style/IEEEtran_v111}
\bibliography{./style/IEEEabrv,./style/RefAbrv,Reference}

\begin{thebibliography}{10}
\providecommand{\url}[1]{#1}
\csname url@rmstyle\endcsname
\providecommand{\newblock}{\relax}
\providecommand{\bibinfo}[2]{#2}
\providecommand\BIBentrySTDinterwordspacing{\spaceskip=0pt\relax}
\providecommand\BIBentryALTinterwordstretchfactor{4}
\providecommand\BIBentryALTinterwordspacing{\spaceskip=\fontdimen2\font plus
\BIBentryALTinterwordstretchfactor\fontdimen3\font minus
  \fontdimen4\font\relax}
\providecommand\BIBforeignlanguage[2]{{%
\expandafter\ifx\csname l@#1\endcsname\relax
\typeout{** WARNING: IEEEtran.bst: No hyphenation pattern has been}%
\typeout{** loaded for the language `#1'. Using the pattern for}%
\typeout{** the default language instead.}%
\else
\language=\csname l@#1\endcsname
\fi
#2}}

\bibitem{ref_Laya}
A.~Laya, L.~Alonso, and J.~Alonso-Zarate, ``Is the random access channel of
  {LTE} and {LTE-A} suitable for {M2M} communications? a survey of
  alternatives,'' \emph{IEEE Commun. Surv. Tut.}, vol.~16, no.~1, pp. 4--16,
  First 2014.

\bibitem{ref_Islam}
M.~T. Islam, A.~e.~M.~Taha, and S.~Akl, ``A survey of access management
  techniques in machine type communications,'' \emph{{IEEE} Commun. Mag.},
  vol.~52, no.~4, pp. 74--81, April 2014.

\bibitem{ref_Osti}
P.~Osti, P.~Lassila, S.~Aalto, A.~Larmo, and T.~Tirronen, ``Analysis of {PDCCH}
  performance for {M2M} traffic in {LTE},'' \emph{{IEEE} Trans. Veh. Technol.},
  vol.~63, no.~9, pp. 4357--4371, Nov 2014.

\bibitem{ref_Jang}
{H. S. Jang, S. M. Kim, K. S. Ko, J. Cha and D. K. Sung}, ``Spatial group based
  random access for {M2M} communications,'' \emph{{IEEE} Commun. Lett.},
  vol.~18, no.~6, pp. 961--964, June 2014.

\bibitem{ref_Jang5}
H.~S. Jang, S.~M. Kim, H.~S. Park, and D.~K. Sung, ``An early preamble
  collision detection scheme based on tagged preambles for cellular {M2M}
  random access,'' \emph{{IEEE} Trans. Veh. Technol.}, vol.~66, no.~7, pp.
  5974--5984, July 2017.

\bibitem{ref_Jang4}
{H. S. Jang, H.-S Park, and D. K. Sung}, ``{A non-orthogonal resource
  allocation scheme in spatial group based random access for cellular M2M
  communications},'' \emph{{IEEE} Trans. Veh. Technol.}, vol.~66, no.~5, pp.
  4496--5000, May 2017.

\bibitem{ref_Liang}
Y.~Liang, X.~Li, J.~Zhang, and Z.~Ding, ``Non-orthogonal random access {(NORA)}
  for {5G} networks,'' \emph{{IEEE} Trans. Wireless Commun.}, vol.~PP, no.~99,
  2017.

\bibitem{ref_Overload2}
T.-M. Lin, C.-H. Lee, J.-P. Cheng, and W.-T. Chen, ``{PRADA: Prioritized Random
  Access With Dynamic Access Barring for MTC in 3GPP LTE-A Networks},''
  \emph{{IEEE} Trans. Veh. Technol.}, vol.~63, no.~5, pp. 2467--2472, June
  2014.

\bibitem{ref_ACB1}
S.~Y. Lien, T.~H. Liau, C.~Y. Kao, and K.~C. Chen, ``{Cooperative access class
  barring for machine-to-machine communications},'' \emph{{IEEE} Trans.
  Wireless Commun.}, vol.~11, no.~1, pp. 27--32, January 2012.

\bibitem{ref_koseoglu2017pricing}
M.~Koseoglu, ``Pricing-based load control of {M2M} traffic for the {LTE-A}
  random access channel,'' \emph{{IEEE} Trans. Commun.}, vol.~65, no.~3, pp.
  1353--1365, 2017.

\bibitem{ref_vilgelm2017latmapa}
M.~Vilgelm, H.~M. G{\"u}rsu, W.~Kellerer, and M.~Reisslein, ``{LATMAPA}:
  Load-adaptive throughput-maximizing preamble allocation for prioritization in
  {5G} random access,'' \emph{IEEE Access}, vol.~5, pp. 1103--1116, 2017.

\bibitem{ref_Ko}
{K. S. Ko, M. J. Kim, K. Y. Bae, D. K. Sung, J. H. Kim, and J. Y. Ahn}, ``A
  novel random access for fixed-location machine-to-machine communications in
  {OFDMA} based systems,'' \emph{{IEEE} Commun. Lett.}, vol.~16, no.~9, pp.
  1428--1431, Sept. 2012.

\bibitem{ref_Wang}
Z.~Wang and V.~W.~S. Wong, ``Optimal access class barring for stationary
  machine type communication devices with timing advance information,''
  \emph{{IEEE} Trans. Wireless Commun.}, vol.~14, no.~10, pp. 5374--5387, Oct
  2015.

\bibitem{ref_LTE}
{S. Sesia, I. Toufik, and M. Baker}, \emph{LTE - The UMTS Long Term Evolution
  From Theory to Practice}.\hskip 1em plus 0.5em minus 0.4em\relax John Wiley
  \& Sons Ltd., 2009.

\bibitem{ref_Jang3}
{H. S. Jang, S. M. Kim, H.-S Park, and D. K. Sung}, ``{Message-embedded random
  access for cellular M2M communications},'' \emph{{IEEE} Commun. Lett.},
  vol.~20, no.~5, pp. 902--905, May 2016.

\bibitem{rice}
S.~O. Rice, ``Mathematical analysis of random noise,'' \emph{Bell System
  Technical Journal}, vol.~24, no.~1, pp. 46--156, 1945.

\end{thebibliography}
\end{document}